\documentclass[%
superscriptaddress,
 amsmath,amssymb,
 aps,
]{revtex4-2}

\usepackage{graphicx}
\usepackage{dcolumn}
\usepackage{bm}
\usepackage{physics}
\usepackage{gensymb}
\usepackage{threeparttable}
\usepackage{tabularx}
\usepackage{hyperref}
\hypersetup{
setpagesize=false,
 bookmarksnumbered=true,%
 bookmarksopen=true,%
 colorlinks=true,%
 linkcolor=blue,
 citecolor=blue,
 urlcolor=blue,
}

\def\mb{\mu_{\rm B}}

\def\tc{T_{\rm C}}

\def\ms{J_{\rm s}}
\def\js{\ms}

\def\ku{K_{\rm u}}

\def\nfb{\rm Nd_2Fe_{14}B}
\def\pfb{\rm Pr_2Fe_{14}B}

\def\npfb{({\rm Nd}_{1-x}{\rm Pr}_{x})_2{\rm Fe}_{14}{\rm B}}

\newcommand{\resec}[1]{\mbox{Sec. \ref{#1}}}
\newcommand{\retable}[1]{\mbox {Table \ref{#1}}}
\newcommand{\ReFig}[1]{\mbox{Figure \ref{#1}}}
\newcommand{\refig}[1]{\mbox{Fig. \ref{#1}}}

\newcolumntype{Y}{>{\centering\arraybackslash}X} 

\begin{document}


\title{Impact of Lattice Distortions on Magnetocrystalline Anisotropy and Magnetization in $\npfb$ Alloys}

\author{Haruki~Okumura}
 \email{h.okumura@aist.go.jp}
\affiliation{Materials DX Research Center, National Institute of Advanced Industrial Science and Technology (AIST), Tsukuba, Ibaraki 305-8568, Japan}

\author{Takashi~Miyake}
\affiliation{Materials DX Research Center, National Institute of Advanced Industrial Science and Technology (AIST), Tsukuba, Ibaraki 305-8568, Japan}

\author{Taro~Fukazawa}
\affiliation{Materials DX Research Center, National Institute of Advanced Industrial Science and Technology (AIST), Tsukuba, Ibaraki 305-8568, Japan}

\author{Noritsugu~Sakuma}
\affiliation{Advanced Material Engineering Division, Toyota Motor Corporation, Susono, Shizuoka 410-1193, Japan}

\author{Yuta~Suzuki}
\affiliation{Advanced Material Engineering Division, Toyota Motor Corporation, Susono, Shizuoka 410-1193, Japan}

\author{Tetsuya~Shoji}
\affiliation{Advanced Material Engineering Division, Toyota Motor Corporation, Susono, Shizuoka 410-1193, Japan}

\author{Hisazumi~Akai}
\affiliation{Department of Precision Science and Technology, Graduate School of Engineering, The University of Osaka, 2-1 Yamadaoka, Suita, Osaka 565-0871, Japan}

\author{Masako~Ogura}
\affiliation{R$^3$ Institute for Newly-Emerging Science Design, The University of Osaka, 1-2 Machikaneyama, Toyonaka, Osaka 560-0043, Japan}

\author{Tetsuya~Fukushima}
\affiliation{Materials DX Research Center, National Institute of Advanced Industrial Science and Technology (AIST), Tsukuba, Ibaraki 305-8568, Japan}

\date{\today}

\begin{abstract}
$\nfb$---a widely used permanent magnet---has magnetocrystalline anisotropy constants that differ between the bulk and interface regions.
This study explores the effects of lattice distortion on the magnetocrystalline anisotropy ($\ku$) and magnetization of $\npfb$.
Nd$_2$Fe$_{14}$B alloys were fabricated; scanning transmission electron microscopy revealed a compressive strain of up to 25\% near grain boundaries.
Using the full-potential Korringa--Kohn--Rostoker method, we calculated the strain dependence of $\ku$, showing that although $\ku$ is 4.2 MJ/m$^3$ under strain-free conditions at 0 K, it becomes negative in regions with 25\% compressive strain.
Additionally, $\pfb$ exhibits a larger $\ku$ than $\nfb$ under undistorted conditions, whereas Pr-rich alloys exhibit a more pronounced reduction in $\ku$ under strain.
These findings highlight the critical influence of lattice distortions on magnetic properties.
The calculated strain-dependent magnetic anisotropy parameters provide valuable inputs for future micromagnetic simulations, aiding the design of advanced magnetic materials.
\end{abstract}

\pacs{Valid PACS appear here}
\maketitle

\section{Introduction}
Permanent-magnet materials are extensively utilized in industrial applications, including motors for electric vehicles and wind turbines. High-performance permanent magnets require high magnetization $\js$, a high magnetocrystalline anisotropy constant $\ku$, and a high Curie temperature $\tc$. The high $\js$ and $\tc$ primarily originate from the $3d$ orbitals of transition metals, whereas the high $\ku$ typically arises from the $4f$ orbitals of lanthanide elements \cite{Miyake2018}. Rare-earth (RE) transition-metal magnetic materials exhibit strong magnetocrystalline anisotropy due to the large spin--orbit coupling from the orbital angular momentum of $4f$ electrons remaining under weak crystal fields. A well-known example is $\nfb$ \cite{Sagawa1984,Croat1984}, which exhibits an anisotropy constant of $K\sim 5$--$6$ MJ/m$^3$, $\js \sim 1.6$ T, and $\tc = 586$ K \cite{Sagawa1985,Koon1985,Givord1985,Yamada1986,Otani1987,Hirosawa1986}.

Dy doping in $\nfb$ has been shown to enhance the magnetic properties, including anisotropy.
However, the geographical distribution and high cost of Dy pose significant challenges for large-scale application.
As a more cost-effective alternative, Pr is being considered for improving the magnetic properties of $\nfb$ alloys because it is more abundant and less expensive.
$\nfb$ and $\pfb$ are closely related materials, with $\pfb$ maintaining out-of-plane anisotropy even at low temperatures, while $\nfb$ undergoes a spin reorientation transition at 135 K \cite{Givord1984,Pique1996,Pedziwiatr1987}.
Because $\npfb$ forms a solid solution, Pr doping reduces the spin reorientation transition temperature of the system \cite{Yang1986,Zhidong1989,Marusi1990,Kim1999,Politova2018,Politova2019}.
The addition of Pr can enhance the magnetic anisotropy constant at low temperatures.

In commercial applications, coercivity---a macroscopic physical property---is also critical. The coercivity can be theoretically predicted using the Stoner--Wohlfarth model; however, there is a substantial discrepancy with experimental values, which is known as Brown's paradox \cite{Brown1945}. Therefore, micromagnetic simulations are often employed to computationally estimate the coercivity \cite{Fukunaga2015,Toga2020,Miyashita2021}. Such simulations involve model calculations that require various input parameters. The determination of these parameters for unknown compounds is challenging. Furthermore, the lattice constants vary from the grain boundary to the bulk, making it essential to examine the dependence of $\ku$ and $\js$ on lattice constants. Although many theoretical studies have been performed \cite{Groot1999,Hrkac2014a,Hrkac2014b,Kubo2014}, first-principles calculations of magnetocrystalline anisotropy remain limited \cite{Yi2017}.

In this study, we examined the effects of lattice distortion on the magnetocrystalline anisotropy and magnetization of $\npfb$. Experimental observations and measurements confirmed the crystal structure and the strain near the interface. The primary objective was to quantitatively analyze how the observed strain influences the anisotropy constant and magnetization through calculations. The findings provide valuable insight into how lattice distortions near the interface can modify the magnetic properties of a material. In the experimental observations, a strain of approximately $-25$\% was observed near the interface, and calculations indicated that the magnetic anisotropy was in-plane for this distortion.

\section{Methods}\label{sec:method}
\subsection{Experimental procedure}\label{method_expt}
(Nd$_{1-x}$Pr$_x$)$_{13.55}$Fe$_{80.54}$B$_{5.91}$ alloys were prepared via arc melting.
They were annealed at 1373 K for 24 h in an Ar atmosphere to homogenize the structure.
Then, they were pulverized and sorted into particles with diameters of $<20$ $\mu$m in an inert atmosphere to create magnetically anisotropic powder.
The crystal structures were examined using powder X-ray diffraction (XRD) with synchrotron radiation sources. Synchrotron XRD patterns were obtained at the BL5S2 beamline of the Aichi Synchrotron Radiation Center using a two-dimensional detector (PILATUS 100K, Dectris).
Each sample was packed into a quartz glass capillary (diameter: 0.3 mm) (hilgenberg) and exposed to X-rays for 20 s to acquire XRD data over a wide 2$\theta$ range.
The wavelength of the X-rays was determined to be 0.619975 \AA\ with a beam energy of 19.9983 keV.
The X-ray wavelength was converted to a Cu K$\alpha$ wavelength of 1.54056 \AA.
Owing to experimental constraints, the 2$\theta$ region between 57.1$\degree$ and 91.5$\degree$ was excluded from measurement and analysis.

Melt-spun ribbons with a nominal composition of Nd$_{13.2}$Fe$_{76.16}$Co$_{4.49}$B$_{5.61}$Ga$_{0.54}$ were prepared.
The crushed ribbons were compacted at 923 K for 3 min in a $\phi$10-mm die under a uniaxial pressure of 400 MPa.
Die upsetting was performed at a strain rate of 0.2 s$^{-1}$ at 1053 K to achieve a 65\% height reduction.
5 mass\% of Nd$_{70}$Cu$_{30}$ alloy was infiltrated into the hot-deformed magnet of 4$\times$4$\times$2 mm$^3$ at 873 K for 165 min using the grain-boundary diffusion technique.
The magnetic properties of the samples were measured using a vibrating-sample magnetometer.

Rietveld refinement of the powder XRD patterns was performed using the GSAS-II software package \cite{Toby2013}.
The background was modeled using the Chebyshev function. Peak profiles were described using a pseudo-Voigt function.
The following parameters were refined: scale factor, background parameters, lattice parameters, atomic positions, isotropic atomic displacement parameters, and profile parameters (Gaussian and Lorentzian coefficients and asymmetry parameter).
The initial structural models for refinement were obtained from \cite{Isnard1995} for $\nfb$ and \cite{Herbst1985} for $\pfb$.

Spherical-aberration-corrected scanning transmission electron microscopy (Cs-STEM) was performed using a JRM-ARM200F instrument (JEOL) to detect distortions at the interfaces of the $\nfb$ grains.
Cs-STEM was performed with high-speed scanning at 2 $\mu$s/pixel to minimize the drift effects of the sample.
Ten images were captured continuously, and the images were accumulated while performing drift correction between measurements.
Kernel density estimation was performed to reduce noise in the STEM images.
STEM observations in the [110] direction make it possible to distinguish between the 4$f$ and 4$g$ sites of RE elements.

\subsection{Calculation method}
The electronic structure can be calculated directly using the Korringa--Kohn--Rostoker (KKR) Green's function method, which is based on multiple-scattering theory \cite{Korringa47,Kohn54}. In this study, we used the full-potential KKR (FPKKR) Green's function method \cite{Ogura2005} to calculate $\ku$ and $\js$ at absolute zero temperature.
The potential is defined within a Voronoi cell, where anisotropic contributions are also considered.
The exchange-correlation functional is based on the generalized gradient approximation using the Perdew--Burke--Ernzerhof parameterization \cite{Perdew1996}.
The coherent potential approximation (CPA) \cite{Soven1970,Shiba1971} is employed to treat the potentials in the disordered phase, allowing electronic-structure calculations for nonstoichiometric compositions without constructing supercells. The KKR method combined with CPA (KKR+CPA) has been applied to RE transition-metal permanent magnets \cite{Fukazawa2018,Fukazawa2019,Harashima2020,Okumura2023,Okumura2025}. 
In the present study, we applied this approach to $\npfb$ for Pr concentrations in increments of $x = 0.1$.

$\nfb$ belongs to space group $P4_2/mnm$ (No.136) and exhibits a tetragonal crystal structure.
Lattice distortions were introduced by maintaining tetragonal symmetry, not orthorhombic symmetry. Electronic-structure calculations were performed by varying only $a$ and $c$.
It should be noted that the present calculation is a bulk calculation with distortion, rather than a slab model.
Regarding site preference for Pr addition, Pr was uniformly added to the crystallographically nonequivalent RE sites, specifically the 4$f$ and 4$g$ sites.
However, it has been reported that in first-principles calculations using supercells, the site preference for Pr addition depends on the Pr concentration \cite{Khan2016}.

The Nd and Pr $4f$ orbitals are treated as open-core, where the $4f$ electrons are considered core-like states, and their occupation numbers are fixed to given values during the self-consistent calculations.
Both elements are trivalent, i.e., Nd$^{3+}$ in the $4f^3$ state and Pr$^{3+}$ in the $4f^2$ state, satisfying Hund's rule.
The magnetic moment of the $4f$ electrons is determined by the Russel--Saunders coupling, with the Land\'{e} $g$-factor and total angular momentum $J$ being $g=\frac{8}{11}$ and $J=9/2$ for Nd$^{3+}$ and $g=\frac{4}{5}$ and $J=4$ for Pr$^{3+}$, respectively.
In other words, the contribution of each $4f$ orbital to the total magnetic moment is 3.27 $\mu_B$/atom for Nd$^{3+}$ and 3.20 $\mu_B$/atom for Pr$^{3+}$.
In the open-core treatment, the $4f$ orbitals of the RE elements remain localized and do not hybridize with other orbitals. They influence the magnetic properties, particularly the magnetic anisotropy, through intra-atomic exchange interactions with Nd $5d$ states.
The spin configuration was initially set such that the spins of the Fe $3d$ electrons and the $5d$ electrons of the RE elements were antiparallel, and then the self-consistent calculation was started.

The magnetocrystalline anisotropy constant $\ku$ is defined as follows: The term associated with spin--orbit interactions, which is proportional to the angular momentum along the quantization axis ($l_zs_z$), is considered. Let $E_{001}$ and $E_{100}$ represent the total energy per unit cell (containing 68 atoms) when the quantization axis is aligned along the [001] and [100] axes, respectively.
Then, $\ku$ is calculated using the formula $\ku=(E_{100}-E_{001})/V$, where $V$ represents the volume of the unit cell.
This calculation considers the spin--orbit coupling contributions that determine the magnetic anisotropy of the system.
According to this definition, a positive value of $\ku$ indicates that the system exhibits uniaxial anisotropy with the easy axis of magnetization along the [001] direction, which corresponds to a lower-energy state when the quantization axis lies along this direction.

\section{Experimental Results}
\subsection{Determination of crystal structures}
\begin{table*}[t!]
\caption{
Internal coordinates of atoms determined via Rietveld analysis with XRD measurements.
}
\label{tb:structrure}
	\begin{tabular*}{\textwidth}{@{\extracolsep{\fill}}ccccccccc}
\hline\hline
&&&&$\nfb$&&&$\pfb$\\
Atom&Wyckoff&Symmetry&x&y&z&x&y&z\\\hline
Nd	&4g	&m.2m  &0.23053&0.76953&0.00000&0.22987&0.77013&0.00000\\
Nd	&4f	&m.2m  &0.35683&0.35683&0.00000&0.35842&0.35842&0.00000\\
Fe	&16k	&1 &0.03746&0.35986&0.32475&0.03684&0.36185&0.32494\\
Fe	&16k	&1 &0.06566&0.27616&0.12945&0.06765&0.27635&0.12814\\
Fe	&8j	&..m   &0.09816&0.09816&0.29526&0.09685&0.09685&0.29565\\
Fe	&8j	&..m   &0.31806&0.31806&0.25326&0.31674&0.31674&0.25535\\
Fe	&4e	&2.mm  &0.00000&0.00000&0.11709&0.00000&0.00000&0.11277\\
Fe	&4c	&2/m.. &0.00000&0.50000&0.00000&0.00000&0.50000&0.00000\\
B	&4f	&m.2m  &0.13850&0.13850&0.00000&0.12560&0.12560&0.00000\\
\hline\hline
\end{tabular*}
\end{table*}

\begin{figure*}[ht!]
\centering
\includegraphics[width=1.0\linewidth]{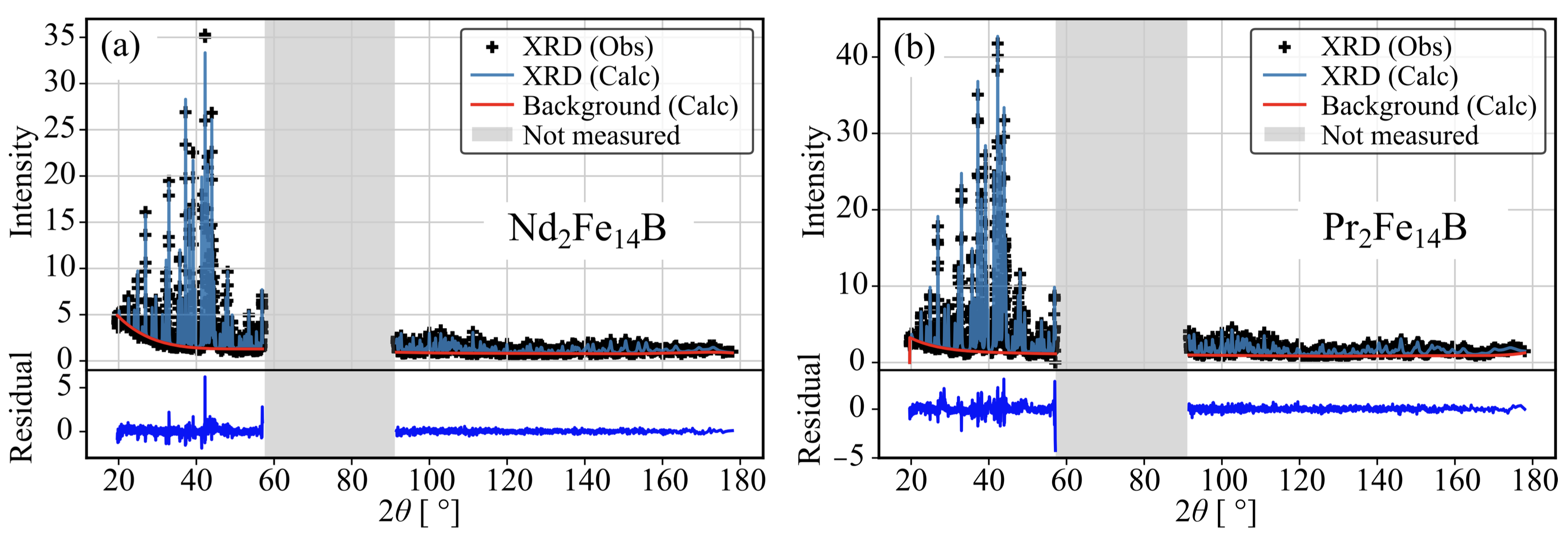}
\caption{Rietveld refinement results of (a) $\nfb$ and (b) $\pfb$, showing the observed XRD data (Obs), fitted pattern (Calc), fitted background, and residual. The $2\theta$ range of 57$\degree$ to 91$\degree$ was not measured, to shorten the data-acquisition time.
}
\label{fig:XRD}
\end{figure*}

\ReFig{fig:XRD} shows the Rietveld refinement results for Nd$_2$Fe$_{14}$B and Pr$_2$Fe$_{14}$B, displaying the experimentally observed XRD data, the fitted patterns, the background, and the residual between the observed and fitted patterns.
Based on these results, the internal coordinates and crystal structures of $\nfb$ and $\pfb$ were determined.
For $\nfb$ ($\pfb$), the experimentally determined lattice constant $a$ was 8.8177 (8.8112) \AA, and the $c/a$ ratio was 1.3851 (1.3902).
The Wyckoff positions for these structures are presented in \retable{tb:structrure}.
The obtained crystal structures closely aligned with several previously reported experimental structures \cite{Herbst1985, Pringle1990, Isnard1995}.
The refinements converged to satisfactory agreement factors, with $R_{\rm wp}$ and goodness-of-fit values of 11.281\% and 0.17 for $\nfb$ and 10.979\% and 0.18 for $\pfb$, respectively.
The $R_{\rm wp}$ values were slightly higher than the typical threshold of $<$10\%, which is attributed to the low peak intensities in the XRD patterns.

\begin{figure}[ht!]
	\centering
	\includegraphics[width=0.7\linewidth]{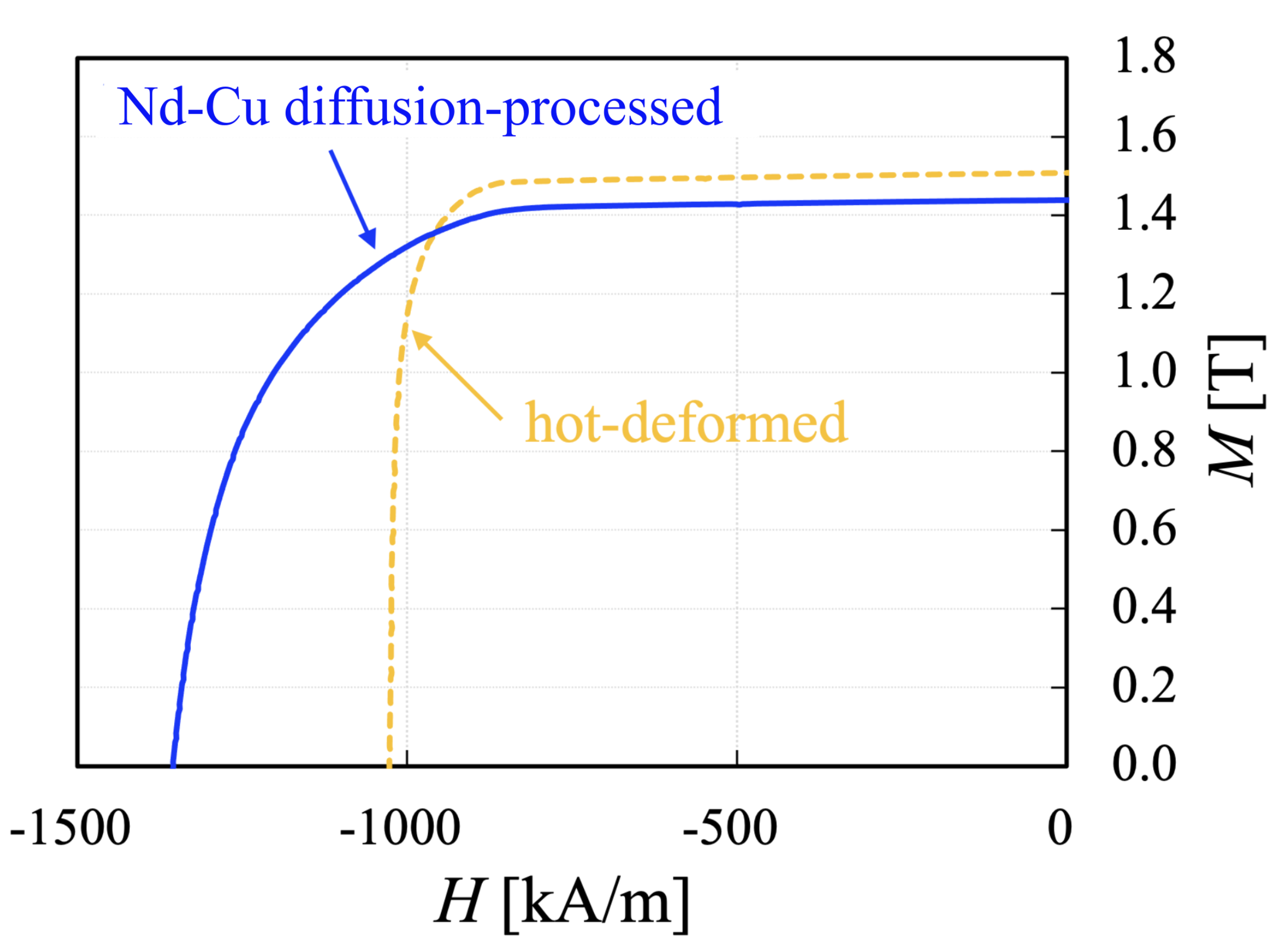}
	\caption{Demagnetization curves of the hot-deformed magnet and the Nd-Cu diffusion-processed magnet.}
	\label{fig:demag_curve}
\end{figure}

\subsection{Measurement of interfacial strain}\label{sec:expt2}
\ReFig{fig:demag_curve} presents the demagnetization curves of the hot-deformed and Nd-Cu diffusion-processed samples.
The coercivity increased significantly from 1027 to 1355 kA/m following the diffusion process, whereas the remanent magnetization decreased slightly from 1.51 to 1.44 T.
The maximum energy product (BH)$_{\rm max}$ was calculated to be 55.7 MGOe for the hot-deformed sample and 50.9 MGOe for the diffusion-processed sample.

The enhancement in coercivity is attributed to the diffusion of NdCu. This diffusion process has two main effects: first, it minimizes the distortion at the interfaces of $\nfb$ grains and restores magnetic anisotropy; second, the diffusion of nonmagnetic NdCu causes magnetic decoupling, reducing the exchange coupling between the main phases and thereby enhancing the coercivity. According to a previous study comparing experiments and calculations \cite{Bance2014}, the size of the interfacial strain region without anisotropy decreases from approximately 2.8 to 1.2 nm after NdCu diffusion. Additionally, VASP calculations predicted \cite{Yi2017} that the amount of strain causing zero anisotropy owing to the equivalent interfacial strain in the a and b directions is approximately 4\%. However, recent reports \cite{Murakami2015, Murakami2016} based on actual strain measurements indicated the presence of small strains ($\pm$1\%) at the grain surfaces of $\nfb$ magnets. Such small observed strains have little effect on the coercivity and cannot explain the enhanced coercivity of the diffusion-processed sample.

\begin{figure*}[t!]
	\centering
	\includegraphics[width=1.0\linewidth]{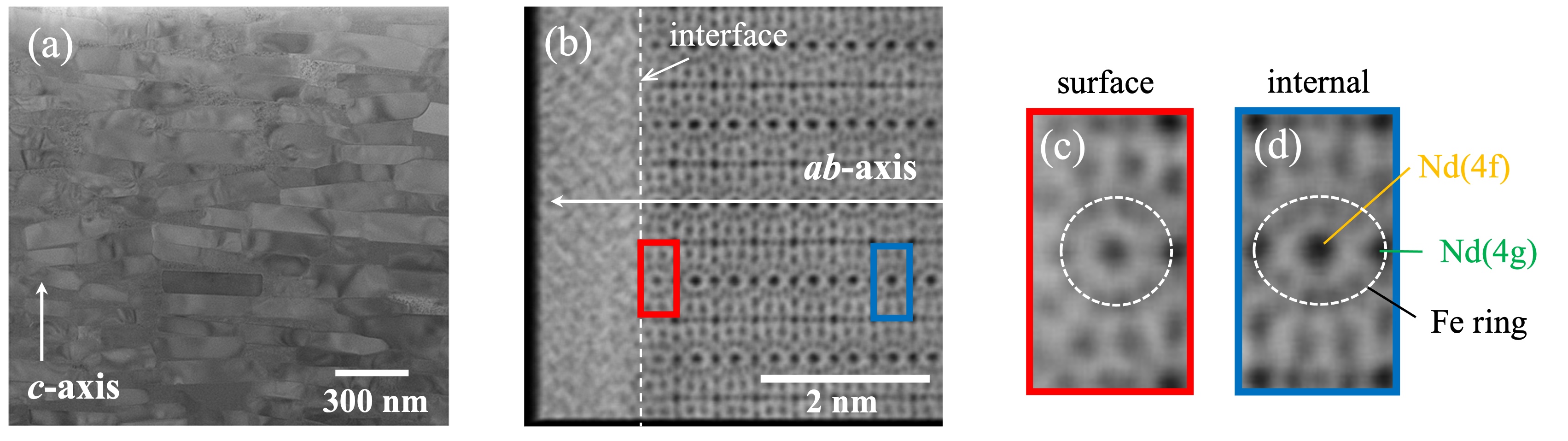}
	\caption{Cs-STEM images of Nd-Cu diffusion-processed magnets.
		(a) Bright-field STEM image showing the in-plane $c$-axis.
		(b) Drift-corrected Cs-STEM image along the $ab$-axis (along the [110] direction).
		(c) Unit-cell images of the surface and (d) internal regions.
	}
	\label{fig:stem}
\end{figure*}

\begin{figure*}[t!]
	\centering
	\includegraphics[width=1.0\linewidth]{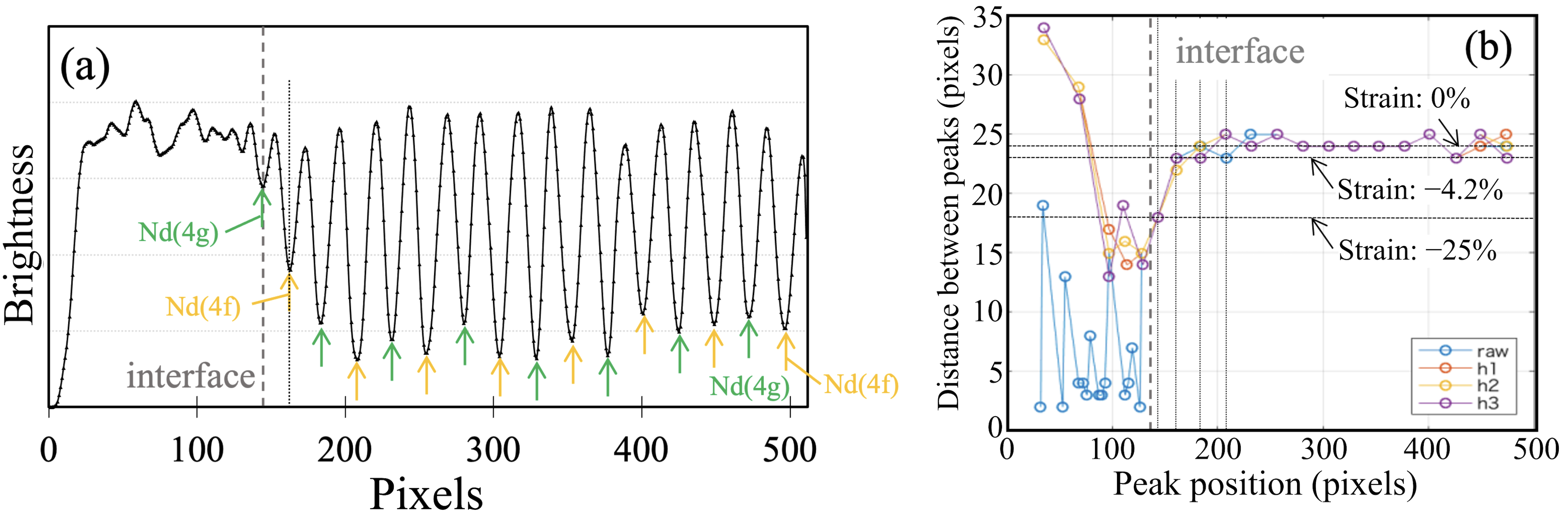}
	\caption{Strain analysis of Nd-Cu diffusion-processed magnets.
		(a) Brightness profile along the white arrow in \refig{fig:stem}.
		The minima correspond to the centers of the Nd $4f$ and $4g$ sites.
		(b) Quantification of interfacial strain based on the differences in distances between adjacent peaks.
		$h$ represents the value of the smoothing parameter in kernel density estimation. A larger $h$ indicates more smoothing.
	}
	\label{fig:strain_analysis}
\end{figure*}

To further examine the strain variation as a function of distance from the interface, Cs-STEM images were obtained, as shown in \refig{fig:stem}.
The sample after grain-boundary diffusion was used because the grain-boundary phase became thicker, reducing the overlap of the main phase in the depth direction during STEM observation.
 Additionally, this facilitated the observation of edge-on interfaces, where the interface between the main phase and grain-boundary phase is perpendicular. Moreover, because hot-deformed magnets contain more grains within the area cut out by the focused ion beam than sintered magnets, it is easier to find a grain along the [110] zone axis.
Nd and Fe atomic columns are distinctly observed in the drift-corrected Cs-STEM image in \refig{fig:stem}(b).
The surface aspect ratio of the Fe ring is smaller than that of the internal region, suggesting compression of the surface Fe ring along the [110] direction ($ab$-axis in \refig{fig:stem}(b)).
The strain was quantified using the results presented in \refig{fig:strain_analysis}.
A line analysis of brightness was conducted along the arrow in \refig{fig:stem}(b), where the local brightness minima correspond to the Nd $4f$ and $4g$ sites.

The pixel variation between the $4f$ and $4g$ sites was further analyzed, as shown in \refig{fig:strain_analysis}(b). To quantify the interfacial strain, the internal strain within the grains was set to 0\%.
The analysis revealed that the compressive strain gradually developed within approximately 1 nm of the surface, reaching a maximum value of 25\% at the grain edge.
\ReFig{fig:strain_analysis} also highlights lattice distortion along the [110] direction, transitioning from the bulk region to the interfacial region.
This systematic evaluation provides insights into the interfacial strain behavior and its influence on the structural properties of $\nfb$ magnets.

Table \ref{tb:strain} presents the distances between the interfaces and the corresponding strain values.
A compressive strain of up to 25\% along the [110] direction ($\delta a$) was observed near the interface.
In regions with such significant lattice distortions, the magnetocrystalline anisotropy constant deviates considerably from the bulk value.
This pronounced compressive strain causes a substantial reduction in the magnetic anisotropy at the interface, leading to a notable decrease in the theoretically estimated coercivity of $\nfb$ magnets.
Importantly, the experimentally measured magnitude and spatial extent of the strain region aligned well with the predictions of several computational studies \cite{Bance2014, Ali2024}.

\begin{table*}[t!]
\caption{
The distance from the interface to the internal region along the [110] direction ($ab$-axis) and the corresponding strain (denoted as $\delta a$) evaluated from Cs-STEM images.
The $\ku$ (FPKKR) was calculated with $c$ fixed and $a$ varied. Previous calculations were taken from Ref.\cite{Yi2017}.
}
\label{tb:strain}
	\begin{tabular*}{\textwidth}{@{\extracolsep{\fill}}cccc}
\hline\hline
Distance from surface & Strain $\delta a$& $\ku$(FPKKR) & $\ku$ \cite{Yi2017}\\
(nm) & (\%) & (MJ/m$^3) $  & (MJ/m$^3$) \\ \hline
0.26 & $-25$   &   $-33$   & $<0$ \\
0.53 & $-8.3$  &   $+1.2$  & $<0$ \\
0.85 & $-4.2$  &   $+3.3$  & $<0$ \\
1.15 &   0     &   $+4.2$   & $\simeq 5$\\
\hline\hline
\end{tabular*}
\end{table*}


\section{Calculation Results}
\subsection{Magnetic properties of $\nfb$ with lattice distortions}
Before presenting the calculation results with distortion, we present the results without distortion.
The calculated $\ku$ was 4.2 MJ/m$^3$, which was close to the experimental value \cite{Sagawa1985,Koon1985,Givord1985,Yamada1986,Otani1987} and the previous calculated value \cite{Yi2017}.
The calculated magnetization was 1.89 T, which was close to the experimental value of 1.85 T at 4.2 K \cite{Hirosawa1986}, and the calculated magnetic moment per formula unit was 38.45 $\mu_\mathrm{B}$/f.u. at 0 K, in good agreement with the experimental values at 4.2 K \cite{Givord1984,Hirosawa1986,Tokuhara1985} and with previous calculations \cite{Yi2017}. In our calculations, the Fe and Nd moments were assumed to adopt a collinear magnetic configuration.

\begin{figure}[b!]
	\centering
	\includegraphics[width=0.7\linewidth]{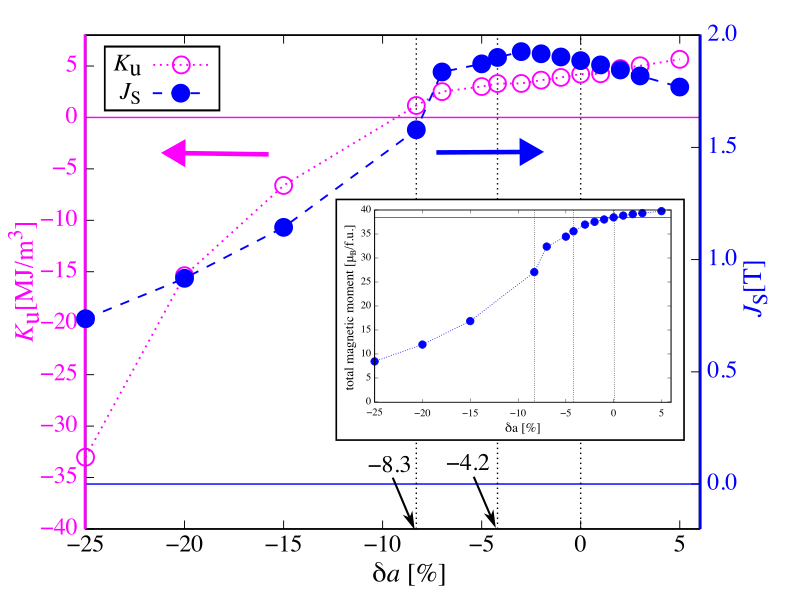}
	\caption{
		The $\delta a$ dependence of $\ku$ and $\js$ for $\nfb$, where the lattice constant $c$ is fixed ($\delta c = 0$).
		The inset shows the $\delta a$ dependence of the total magnetic moment, with the unit of $\mb$ per formula unit.
	}
	\label{fig:calc1}
\end{figure}

As $\delta a$ varies from 5 to $-25$\%, $\ku$ decreases monotonically.
The reduction caused by the hybridization between Fe $3d$ and Nd $5d$ orbitals arises from the rearrangement of the charge distribution around the Nd ions.
This rearrangement modifies the crystal field acting on the $4f$ states.
This mechanism is analogous to that described in Ref. \cite{Toga2015}, where the Nd-O hybridization induces a similar rearrangement of charge density and consequently alters the crystal field parameter.
$\js$ increases slightly owing to compression and then begins to decrease.
Note that if we evaluate only the total magnetic moment, as shown in the inset of \refig{fig:calc1}, the magnetic moment decreases monotonically owing to compression along the [110] direction.
Hence, in the region of increasing magnetization, the effect of decreasing the magnetic moment is greater than that of decreasing the cell volume.

\begin{figure}[ht!]
	\centering
	\includegraphics[width=0.8\linewidth]{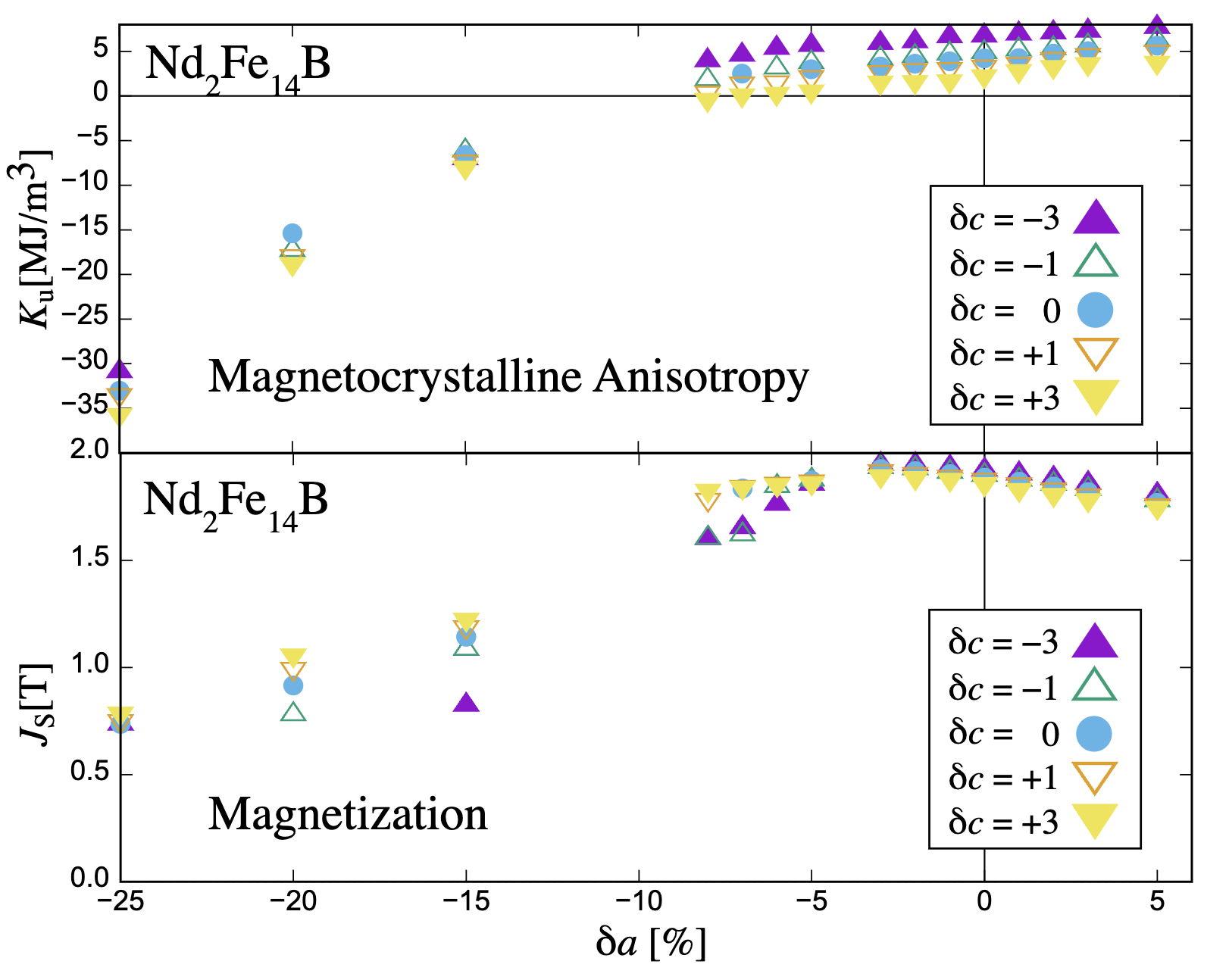}
	\caption{
		Lattice-constant dependence of $\ku$ and $\js$ calculated using FPKKR, based on the bulk crystal structure.
	}
	\label{fig:calc2}
\end{figure}

Next, we present the results for $\nfb$ under the strain applied in the $ab$-plane and along the $c$-axis.
\ReFig{fig:calc2} illustrates the lattice-constant dependence of $\ku$ and $\js$, where $\delta a$ and $\delta c$ represent the amounts of strain along the [110] and [001] directions, respectively.
Focusing on the case in which the strain is induced only along the [001] direction ($\delta a=0$), $\ku$ decreases as the $c$-axis is extended.
Conversely, when $\delta c$ is fixed, $\ku$ decreases with compression along the [110] direction.
These results indicate that $\ku$ decreases as the $c/a$ ratio increases---a trend consistent with previous calculations \cite{Yi2017}.

This phenomenon was explained in \cite{Yi2017,Toga2015} as follows:
The $5d$ electron cloud of the Nd atoms extends toward Fe/B atoms through hybridization.
This extension causes the $4f$ electron cloud to rearrange toward a direction where the $5d$ electron cloud is less extended to minimize the Coulomb repulsion.
Consequently, the orbital magnetic moments of the Nd ions' $4f$ electrons, which have oblate orbitals, align along the [110] direction.
In other words, as the $c/a$ ratio increases, $\ku$ decreases because the easy axis tends to shift in the in-plane direction.

The calculations in \cite{Yi2017} demonstrated that $\ku$ becomes negative when the $ab$-plane is compressed by 4\%.
In contrast, our calculations indicate that $\ku$ remains positive until $\delta a \sim -10$\%.
At the extreme compression level of $\delta a = -25$\%, as observed experimentally in \resec{sec:expt2}, $\ku$ reaches a significantly negative value in the FPKKR calculations.
However, in this region, the value of the magnetization $\ms$ also decreases significantly.
Thus, the impact of strain in the grain-boundary region on the overall coercivity is unlikely to be substantial.
This observation suggests that while the local strain can modify magnetic properties, such as $\ku$, its contribution to the macroscopic coercivity may be mitigated by the simultaneous reduction in $\js$.

\subsection{Pr dependence of magnetocrystalline anisotropy}
\begin{figure}[b!]
	\centering
	\includegraphics[width=1.0\linewidth]{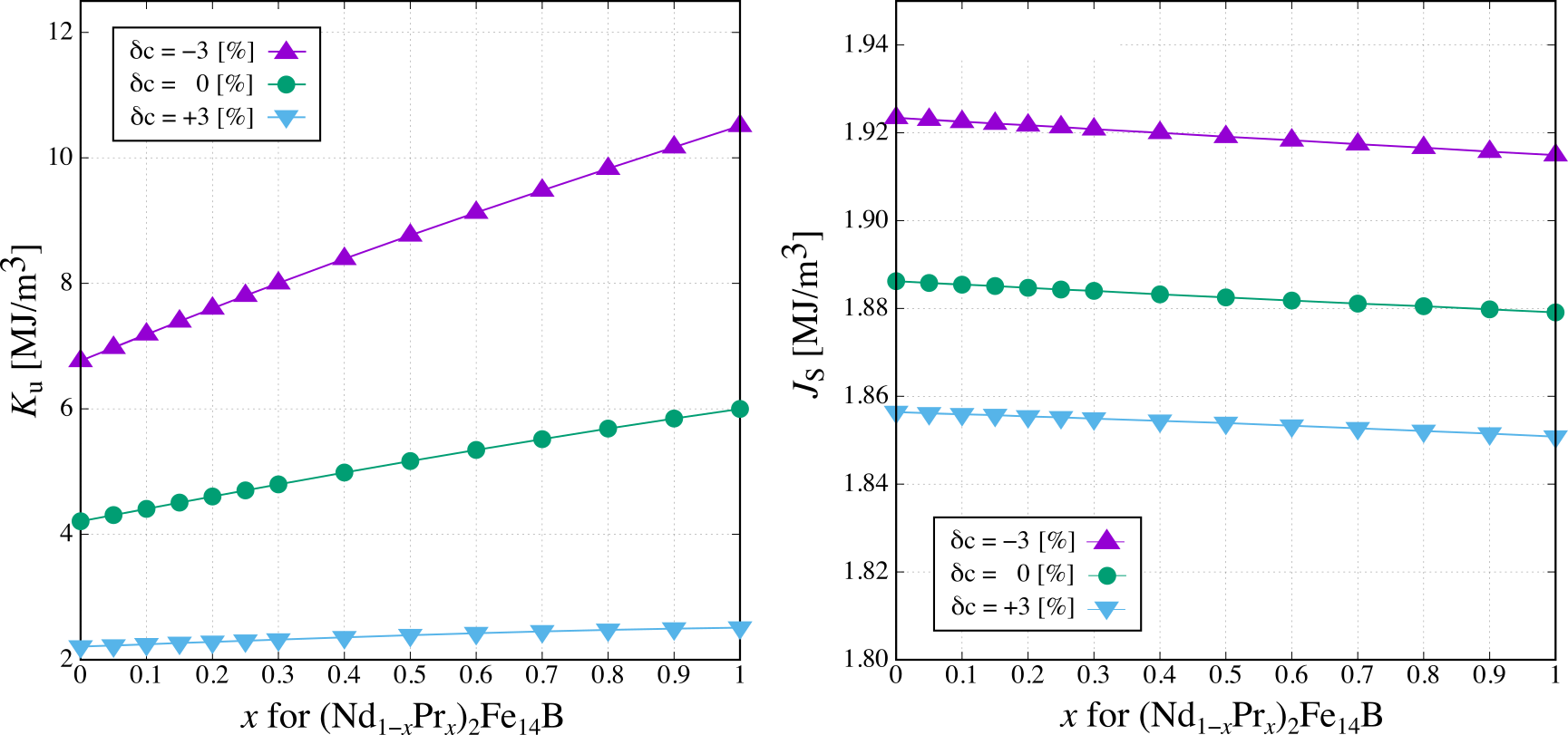}
	\caption{
		Calculated values of the magnetic anisotropy constant and magnetization with respect to the Pr concentration, where $\delta c=0,\pm 3$.
	}
	\label{fig:calc3}
\end{figure}

\refig{fig:calc3} presents the calculated dependence of $\ku$ and $\js$ on the Pr concentration in $\npfb$, showing the results for $\delta c = -3\%, 0\%, and 3$\%.
In the FPKKR calculations, $\ku$ of $\pfb$ is 6.0 MJ/m$^3$, which is larger than $\ku = 4.2$ MJ/m$^3$ for $\nfb$.
These results are consistent with the experimental results \cite{Kim1999} at 4.2 K, indicating that the anisotropy constants $K_1 + K_2$ of $\pfb$ are larger than those of $\nfb$.
The results for $\delta c = 0$\% indicate that $\ku$ varies almost linearly with the Pr concentration, suggesting that the magnetic anisotropy increases with Pr addition at low temperatures.

Meanwhile, $\js$ is slightly lower in $\pfb$ than that in $\nfb$.
This difference is attributed to the smaller effective magnetic moment of Pr$^{3+}$ (3.20 $\mu_B$) compared with that of Nd$^{3+}$ (3.27 $\mu_B$).
Notably, the value of $\js$ (or the total magnetic moment), excluding the contribution of $4f$ electrons to the magnetic moment, is identical for $\nfb$ and $\pfb$, at 1.57 T (31.9 $\mu_B$/f.u.), with a large magnetic moment predominantly arising from Fe.

Next, we consider the case where $\delta c \neq 0$.
Under these conditions, the absolute value of $\js$ varies with $\delta c$; however, its dependence on the Pr concentration remains independent of $\delta c$.
This is because the magnetic moment of Fe $3d$ electrons predominantly determines the value of $\js$, and the contribution of the RE $4f$ electrons is negligible.

In contrast, for $\ku$, the dependence on the Pr concentration is significantly affected by the lattice constants.
This difference arises because $\ku$ is primarily determined by the contribution of RE $4f$ electrons rather than Fe $3d$ electrons.
Specifically, it is influenced by the differences in the charge distributions of the Pr$^{3+}$ and Nd$^{3+}$ $4f$ electrons.
Within the framework of crystal field theory, the deviation of the $4f$ electron charge distribution from spherical symmetry is quantified by the Stevens factor $\alpha_J$.
The $\alpha_J$ values for the Pr$^{3+}$ and Nd$^{3+}$ ground multiplets are $-0.02$ and $-0.006$, respectively.
Both $\alpha_J < 0$ values indicate that the $4f$ electron charge distribution is oblate, but the larger absolute value of $\alpha_J$ for Pr$^{3+}$ implies a greater deviation from spherical symmetry.
Compression along the $c$-axis induces a rearrangement of the charge distribution due to hybridization between Fe $3d$ and RE $5d$, extending the RE $5d$ electron cloud along the $c$-axis.
The $4f$ electrons of Pr$^{3+}$ are more strongly affected by the Coulomb repulsion with the $5d$ electrons than those of Nd$^{3+}$.
As a result, $\ku$ is more sensitive to lattice distortions in Pr-rich systems than in Nd-rich alloys.

\begin{figure}[ht!]
\centering
\includegraphics[width=\linewidth]{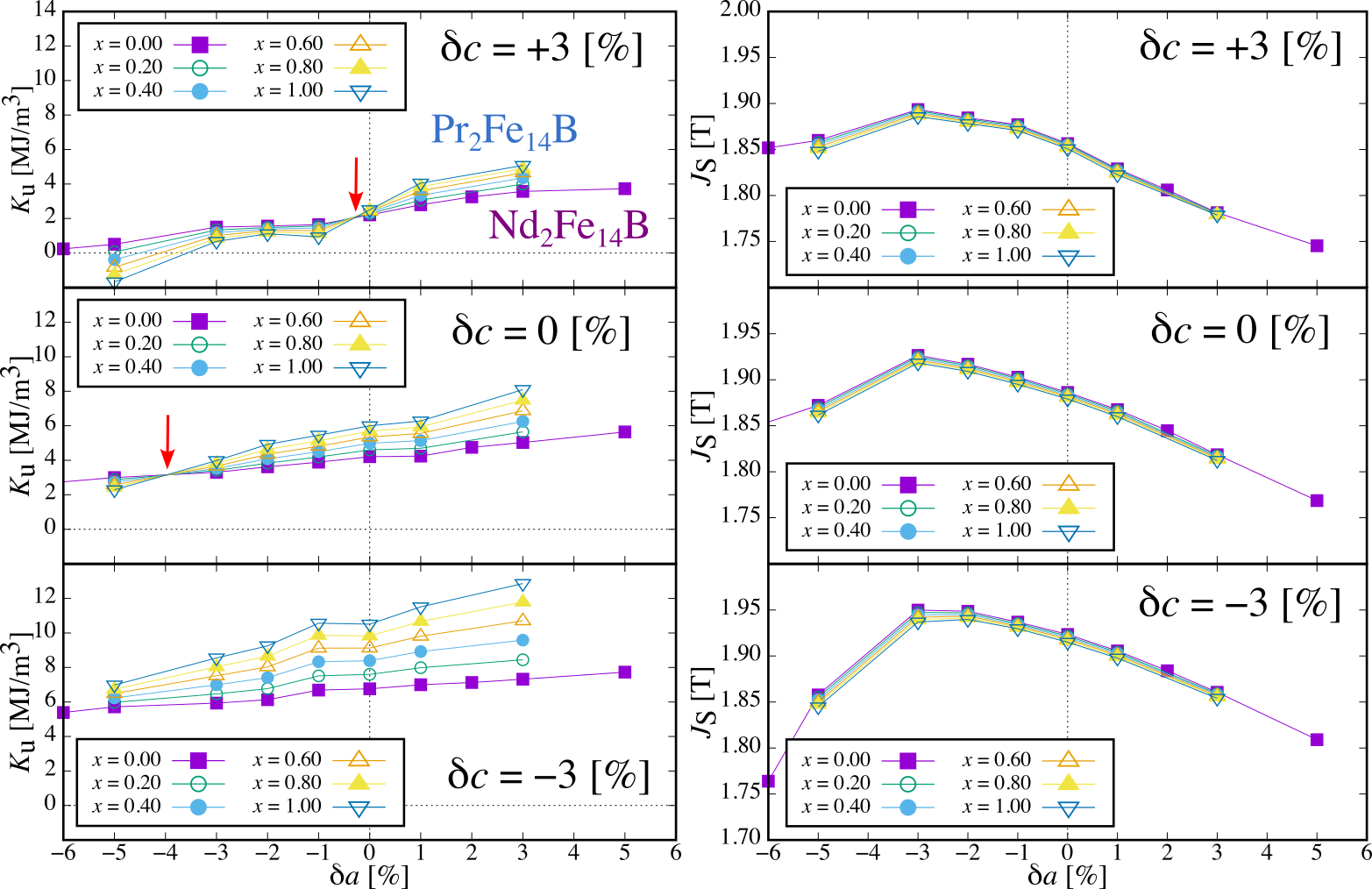}
\caption{
Calculated values of the magnetic anisotropy constant and magnetization of $\npfb$ under compression along the [110] direction, where $\delta c=0,\pm 3$.
The red arrows indicate that the magnetic anisotropy constants of $\nfb$ and $\pfb$ are equal.
}
\label{fig:calc4}
\end{figure}

\ReFig{fig:calc4} shows the $\delta a$ dependence of $\ku$ and $\js$ for various $\delta c$ values, with $x = 0.0, 0.2, 0.4, 0.6, 0.8, 1.0$ for $\npfb$.
For $\delta c = 0$, the $\ku$ of $\pfb$ decreases under compression along the [110] direction, similar to $\nfb$.
This behavior is attributed to the negative sign of $\alpha_J$ for both Pr$^{3+}$ and Nd$^{3+}$ ions.
A larger absolute value of $\alpha_J$ for Pr$^{3+}$ results in a more pronounced decrease in $\ku$ under compression along the [110] direction for Pr-rich systems.
Consequently, the $\ku$ values of $\pfb$ and $\nfb$ converge at approximately $\delta a = -4$\%.
Further compression in the [110] direction beyond this point results in smaller $\ku$ values for Pr-rich systems.
Regarding the magnetization values, the $\delta a$ dependence of $\js$ remains consistent across different Pr concentrations.
However, for the same lattice constant, $\js$ is smaller in Pr-rich alloys because of the smaller effective magnetic moment.

Next, we examine the calculations for various $\delta c$ values.
For $\delta c = +3$\%, even slight compression along the [110] direction leads to a more significant decrease in $\ku$ for $\pfb$ than for $\nfb$.
The $\ku$ value for $\pfb$ becomes negative at approximately $\delta a = -3.5$\%.
Thus, for large $c/a$ ratios and high Pr concentrations, $\ku$ decreases more significantly.
This is attributed to the tendency of the $5d$ electron cloud to expand in the $ab$-plane and the $4f$ electron cloud to elongate along the $c$-axis in crystal structures with large $c/a$ ratios.
For $\delta c = -3$\%, the reduction in $\ku$ with compression along the [110] direction is somewhat suppressed.
Up to $\delta a = -5$\%, the $\ku$ of $\pfb$ remains larger than that of $\nfb$.
Regarding the magnetization values, the main difference from $\delta c$ lies in the absolute values.

As $\delta a$ is further reduced, the $\ku$ of $\pfb$ decreases more significantly than that of $\nfb$.
Observations at the $\nfb$ interface indicate that $\delta a$ reaches approximately $-25$\% in the interfacial region.
Calculations with Pr addition suggest that while $\pfb$ maintains a $\js$ similar to that of $\nfb$, its reduction in $\ku$ is more pronounced.
Consequently, the coercivity may decrease because of the formation of Pr near the interface, in contrast to $\nfb$.
However, experiments involving Pr-doped $\nfb$ powder revealed an increase in the coercivity at Pr concentrations of $x = 0.00, 0.24, 0.40$ at room temperature \cite{Morimoto2016}.
The authors attribute this increase in coercivity to the selective diffusion of Pr into the grain boundaries, where it forms a magnetic phase.
This magnetic phase enhances the coercivity by pinning the magnetic domain walls \cite{Sepehri-Amin2010}.
In addition, our first-principles calculations suggest that Pr substitution enhances the magnetic anisotropy constant in the bulk phase, indicating that the increased coercivity may originate from the main phase.
To theoretically calculate coercivity, more realistic micromagnetic simulations are required.
This study provides parameters for such simulations.

\section{Summary}
This study examined the effects of lattice distortion on the magnetocrystalline anisotropy and magnetization of $\nfb$. STEM measurements revealed a compressive strain of up to 25\% at the grain boundaries, and full-potential KKR calculations with CPA were performed using the corresponding lattice constants. The results indicated that the compressive strain along the [110] direction drives $\ku$ negatively, with calculations predicting a large negative value in the 25\%-strain region, as observed experimentally. The $\ku$ of $\pfb$ exceeds that of $\nfb$ without distortion, which is consistent with experiments conducted at low temperatures \cite{Kim1999}. In Pr-rich alloys, the reduction in $\ku$ under [110] compression is more pronounced, suggesting a possible reduction in coercivity at the interfaces, although our calculations do not reproduce previous reports of coercivity enhancement with Pr addition \cite{Morimoto2016}, likely because of the grain-boundary magnetic phases. The calculated dependence of the magnetic properties on the lattice constants provides essential input for future micromagnetic simulations, bridging theory and experiment.

\section*{Acknowledgments}
The authors greatly acknowledge the “Program for Promoting Researches on the Supercomputer Fugaku” (Computational Research on Materials with Better Functions and Durability Toward Sustainable Development, JPMXP1020230325)
and the "Data Creation and Utilization Type Material Research and Development Project (Digital Transformation Initiative Center for Magnetic Materials)" (Grant No. JPMXP1122715503).
They used the computational resources of the supercomputer Fugaku provided by the RIKEN Center for Computational Science (Project ID: hp250227).
The computations in this study were performed using the facilities of the Supercomputer Center, Institute for Solid State Physics, University of Tokyo.
We would like to thank Editage (www.editage.jp) for English language editing.




\bibliography{bib/shorttitles, bib/citation}
\bibliographystyle{apsrev4-1}
\end{document}